\documentclass[12pt]{article}
\def\beq{\begin{equation}}
\def\eeq{\end{equation}}
\def\bea{\begin{eqnarray}}
\def\eea{\end{eqnarray}}
\usepackage{graphicx}
\usepackage{color}
\usepackage{colortbl}
\usepackage{array}

\catcode`\@=11
\def\lsim{\mathrel{\mathpalette\@versim<}}
\def\gsim{\mathrel{\mathpalette\@versim>}}
\def\@versim#1#2{\vcenter{\offinterlineskip
\ialign{$\m@th#1\hfil##\hfil$\crcr#2\crcr\sim\crcr } }}
\catcode`\@=12
\usepackage{axodraw}

\def\vevof#1{\left\langle#1\right\rangle}
\def\mut{\xi}
\def\rat{r}

\catcode`@=12
\begin{document}

\thispagestyle{empty}
\begin{flushright}
UCRHEP-T515\\
February 2012\
\end{flushright}
\vspace{0.3in}
\begin{center}
{\LARGE \bf Vector-Boson-Induced Neutrino Mass\\}
\vspace{1.5in}
{\bf Ernest Ma$^{a,b}$ and Jose Wudka$^a$\\}
\vspace{0.2in}
{\sl $^a$ Department of Physics and Astronomy, University of California,\\
Riverside, California 92521, USA\\}
\vspace{0.1in}
{\sl $^b$ Institute of Advanced Studies, Nanyang Technological University,
Singapore\\}
\end{center}
\vspace{1.5in}
\begin{abstract}\
One-loop radiative Majorana neutrino masses through the exchange of scalars
have been considered for many years. We show for the first time how such a
one-loop mass is also possible through the exchange of vector gauge bosons.
It is based on a simple variation of a recently proposed $SU(2)_N$ extension
of the standard model, where a vector boson is a candidate for the dark
matter of the Universe. \end{abstract}

\newpage
\baselineskip 24pt

The unique dimension-five operator for Majorana neutrino mass in the
standard model, i.e~\cite{w79}
\begin{equation}
{\cal L}_5 = {f_{ij} \over \Lambda} (\nu_i \phi_0 - l_i \phi^+) (\nu_j
\phi_0 - l_j \phi^+) + H.c.,
\end{equation}
where $(\nu_i,l_i)$ is the usual lepton doublet with $i=e,\mu,\tau$, and
$(\phi^+,\phi^0)$ is the Higgs doublet, is realized at tree level in three
ways~\cite{m98}, through the exchange of a fermion singlet (Type I), a
scalar triplet (Type II), or a fermion triplet (Type III). It may also be
realized in one loop in three ways~\cite{m98} through the exchange of a
scalar and a fermion.  The exchange of two $W$ bosons also contributes in
two loops~\cite{bm88} if one of the neutrinos already has a nonzero Majorana
mass.  Here we show for the first time how a one-loop neutrino mass may also
be generated through the exchange of vector gauge bosons. It is based on a
simple variation of a recently proposed $SU(2)_N$
extension~\cite{dm11,bdmw11} of the standard model, where a vector boson is
a candidate for the dark matter of the Universe~\cite{bhs05}.

The $SU(2)_N$ gauge group contains no component of the electric charge
operator.  It is a subgroup in the decomposition of $E_6$ to $SU(3)_C \times
SU(2)_L \times U(1)_Y \times SU(2)_N$.  It was first discovered~\cite{lr86}
from the consideration of superstring-inspired $E_6$ models~\cite{hr89}.
Its relevance for dark matter was first pointed out recently~\cite{dm11} in
a nonsupersymmetric model with the imposition of a global U(1) symmetry $S$,
such that a generalized lepton number $L = S + T_{3N}$ remains unbroken
after $SU(2)_N$ is completely broken spontaneously.  In that case, one of
the vector gauge bosons $X_1$ of $SU(2)_N$ becomes a good candidate for dark
matter~\cite{bdmw11}.

Under $SU(3)_C \times SU(2)_L \times U(1)_Y \times SU(2)_N \times S$, where
$Q = T_{3L} + Y$ is the electric charge and $L = S + T_{3N}$ is the
generalized lepton number, the fermions of this nonsupersymmetric model are
given by~\cite{dm11}
\begin{eqnarray}
&& \pmatrix{u \cr d} \sim (3,2,1/6,1;0), ~~~ u^c \sim (3^*,1,-2/3,1;0), \\
&& (h^c,d^c) \sim (3^*,1,1/3,2;-1/2), ~~~ h \sim (3,1,-1/3,1;1), \\ &&
\pmatrix{N & \nu \cr E & e} \sim (1,2,-1/2,2;1/2), ~~~ \pmatrix{E^c \cr N^c}
\sim (1,2,1/2,1;0), \\ && e^c \sim (1,1,1,1;-1), ~~~ (\nu^c,n^c) \sim
(1,1,0,2;-1/2),
\end{eqnarray}
where all fields are left-handed.  The $SU(2)_L$ doublet assignments are
vertical with $T_{3L} = \pm 1/2$ for the upper (lower) entries. The
$SU(2)_N$ doublet assignments are horizontal with $T_{3N} = \pm 1/2$ for the
right (left) entries.  There are three copies of the above to accommodate
the known three generations of quarks and leptons, together with their
exotic counterparts.  It is easy to check that all gauge anomalies are
canceled. The extra global U(1) symmetry $S$ is imposed so that $(-1)^L$,
where $L = S + T_{3N}$, is conserved, even though $SU(2)_N$ is completely
broken.  The imposition of $S$ in this case amounts to a generalized lepton
number.  Such a procedure is very commonplace in model building.  For
example, it is used to avoid rapid proton decay
in supersymmetric extensions of the standard model.

The Higgs sector consists of one bidoublet, two doublets, and one $SU(2)_R$
triplet:
\begin{eqnarray}
&& \pmatrix{\phi_1^0 & \phi_3^0 \cr \phi_1^- & \phi_3^-} \sim
(1,2,-1/2,2;1/2), ~~~ \pmatrix{\phi_2^+ \cr \phi_2^0} \sim (1,2,1/2,1;0), \\
&& (\chi_1^0,\chi_2^0) \sim (1,1,0,2;-1/2), ~~~ \pmatrix{\Delta^0_2/\sqrt{2}
& \Delta^0_3 \cr \Delta^0_1 & -\Delta^0_2/\sqrt{2}} \sim (1,1,0,3;1).
\end{eqnarray}
In the following, we differ from the previous proposal by imposing an extra
$Z_2$ symmetry under which $u^c$ and $\phi_2$ are odd, but all other
particles are even.  The allowed Yukawa couplings are thus
\begin{eqnarray}
&& (d \phi_1^0 - u \phi_1^-) d^c - (d \phi_3^0 - u \phi_3^-) h^c, ~~~ (u
\phi_2^0 - d \phi_2^+) u^c, ~~~ (h^c \chi_2^0 - d^c \chi_1^0) h, \\ &&  (E
E^c - N N^c) \chi_2^0 - (e E^c - \nu N^c) \chi_1^0, ~~~ (E^c \phi_1^- - N^c
\phi_1^0) n^c - (E^c \phi_3^- - N^c \phi_3^0) \nu^c, \\ && (N \phi_3^- - \nu
\phi_1^- - E \phi_3^0 + e \phi_1^0) e^c, ~~~ n^c n^c \Delta_1^0 + (n^c \nu^c
+ \nu^c n^c) \Delta_2^0/\sqrt{2} - \nu^c \nu^c \Delta_3^0.
\end{eqnarray}
There are five nonzero vacuum expectation values: $\langle \phi_1^0 \rangle
= v_1$, $\langle \phi_2^0 \rangle = v_2$, $\langle \Delta_1^0 \rangle =
u_1$, and $\langle \chi_2^0 \rangle = u_2$,  corresponding to scalar fields
with $L=0$, as well as $\langle \Delta_3^0 \rangle = u_3$, which breaks $L$
to $(-1)^L$.   Thus $m_d, m_e$ come from $v_1$, and $m_u$ comes from $v_2$,
whereas $m_h, m_E (= -m_{N N^c})$ come from $u_2$, the $N^c n^c$ mass terms
from $v_1$, and $n^c$, $\nu^c$ Majorana masses from $u_1$ and $u_3$. The
scalar fields $\phi_3^{0,-}$ and $\Delta_2^0$ have $L=1$, whereas $\chi_1^0$
has $L=-1$ and $\Delta_3^0$ has $L=2$. The imposed $Z_2$ symmetry is broken
spontaneously by $v_2$ and softly by the trilinear scalar term $\phi_2^0
(\phi_1^0 \chi_2^0 - \phi_3^0 \chi_1^0) - \phi_2^+ (\phi_1^- \chi_2^0 -
\phi_3^- \chi_1^0)$.

There are five neutral fermions per family.  The $3 \times 3$ mass matrix
spanning the $L=0$ fermions $(N,N^c,n^c)$ is given by
\begin{equation}
{\cal M}_N = \pmatrix{0 & -m_E & 0 \cr -m_E & 0 & m_1 \cr 0 & m_1 &
m_{n^c}},
\end{equation}
where $m_E$ comes from $u_2$, $m_1$ from $v_1$, and $m_{n^c}$ from $u_1$.
The $2 \times 2$ mass matrix for the $L=\pm 1$ fermions $(\nu,\nu^c)$ at
tree level is given by \begin{equation}
{\cal M}_\nu = \pmatrix{0 & 0 \cr  0 & m_{\nu^c}},
\end{equation}
However, a radiative Dirac mass linking $\nu$ and $\nu^c$ will appear in one
loop.  Together with the large $\nu^c$ mass, a small Majorana mass for $\nu$
will thus be generated through the usual seesaw mechanism.

Even though this model is nonsupersymmetric, $R$ parity as defined in the
same way as in supersymmetry, i.e. $R \equiv (-)^{3B+L+2j}$, still holds, so
that the usual quarks and leptons (including $\nu^c$) have even $R$, whereas
$h, h^c, (N,E), (E^c,N^c)$, and $n^c$ have odd $R$.  As for the scalars,
$(\phi_1^0,\phi_1^-)$, $(\phi_2^+,\phi_2^0)$, $\chi_2^0$, $\Delta_1^0$, and
$\Delta_3^0$ have even $R$, whereas $(\phi_3^0,\phi_3^-)$, $\chi_1^0$, and
$\Delta_2^0$ have odd $R$. After spontaneous symmetry breaking, the gauge
boson mass eigenstates are
the $W^\pm$ and  $X_{1,2}$, and 3 neutral $R$-even vector bosons
\begin{eqnarray}
A  &=& c_W B + s_W W_3, \cr
Z  &=&  s_W c_\alpha B  - c_W c_\alpha W_3  + s_\alpha X_3, \cr
Z' &=& -s_W s_\alpha B  + c_W s_\alpha W_3  + c_\alpha X_3,
\eea
where $B$ is the $U(1)_Y$ gauge vector field, $c_W = \cos \theta_W$, $
c_\alpha = \cos \alpha$, etc.; with masses
\bea
m_W^2 &=& {1 \over 2} g_2^2 (v_1^2 + v_2^2), \cr
m_{X_{1,2}}^2 &=& {1 \over 2} g_N^2 [u_2^2 + v_1^2 + 2(u_1 \mp u_3)^2], \cr
m_A^2 &=& 0, \cr
m_Z &=& a_+ - \sqrt{a_-^2 + b^2}, \cr
m_{Z'} &=& a_+ + \sqrt{a_-^2 + b^2}, \eea
where
\bea
a_\pm &=&{1 \over 4} g_N^2 \left( 4 u_1^2 + u_2^2 + 4 u_3^2 + v_1^2 \right)
\pm {1 \over 4} \left(g_1^2 + g_2^2 \right) (v_1^2 + v_2^2), \cr
b &=& {1 \over 2} g_N \sqrt{g_1^2 + g_2^2} \, v_1^2, \cr
\tan 2\alpha &=& \frac b{a_-}.
\eea
In the limit $ u_i \gg v_j $ for all $i,j$, $ \alpha \simeq0 $ and $ Z'
\simeq X_3 $.

Whereas the usual gauge bosons have even $R$, two of the $SU(2)_N$ gauge
bosons $X_{1,2}$ have odd $R$ and $X_3 (\simeq Z')$ has even $R$.  Assuming
that $X_1$ is the lightest particle of odd $R$, it becomes a good candidate 
for dark matter~\cite{bdmw11}.  There is also $Z-Z'$ mixing in this model, 
given by $-(\sqrt{g_1^2+g_2^2}/g_N)[v_1^2/(u_2^2+4u_1^2+4u_3^2)]$.  This is
constrained by precision electroweak data to be less than a few times
$10^{-4}$~\cite{pdg10}.  If $m_{Z'} \sim 1$ TeV, then $v_1$ should be less
than about 10 GeV.  Now $m_b$ comes from $v_1$, so this model implies that
$\tan \beta = v_2/v_1$ is large and the Yukawa coupling of $b b^c \phi_1^0$
is enhanced.

Neglecting the contributions of $v_{1,2}$ compared to $u_{1,2,3}$, the
would-be Goldstone bosons for the longitudinal components of $X_{1,2}$ are
given by ($c_1 = \cos \gamma_1 $, etc.)
\begin{eqnarray}
G_1 &=& c_1 \chi_{1I} - s_1 \Delta_{2I} = {u_2 \chi_{1I} - \sqrt{2}(u_1-u_3)
\Delta_{2I} \over \sqrt{u_2^2 + 2(u_1-u_3)^2}}, \\ G_2 &=& c_2 \chi_{1R} +
s_2 \Delta_{2R} = {u_2 \chi_{1R} + \sqrt{2}(u_1+u_3) \Delta_{2R} \over
\sqrt{u_2^2 + 2(u_1+u_3)^2}},
\end{eqnarray}
and the orthogonal linear combinations are physical scalars $H_{1,2}$. The
one-loop radiative neutrino mass is then obtained from the exchange of
$X_{1,2}$, $G_{1,2}$, and $H_{1,2}$ with the $(N,N^c,n^c)$ fermions, as
shown in Figs.~1 and 2.  Note that both diagrams vanish if $m_1=0$ in
Eq.~(11).  As it is, they are finite and calculable.
\begin{figure}[htb]
\begin{center}
\begin{picture}(480,120)(0,0)
\ArrowLine(120,10)(160,10)
\ArrowLine(160,10)(240,10)
\ArrowLine(320,10)(240,10)
\ArrowLine(360,10)(320,10)
\PhotonArc(240,10)(80,0,180)49

\Text(140,0)[]{\large $\nu$}
\Text(340,0)[]{\large $\nu^c$}
\Text(200,0)[]{\large $N$}
\Text(280,0)[]{\large $n^c$}
\Text(240,105)[]{\large $X_{1,2}$}
\Text(240,10)[]{\Large $\times$}

\end{picture}
\end{center}
\caption{One-loop generation of neutrino mass from the vector bosons 
$X_{1,2}$.}
\end{figure}
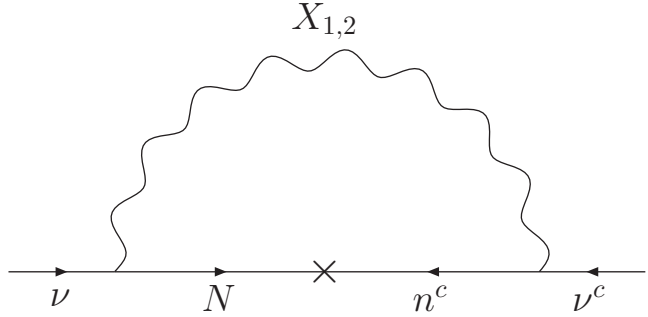
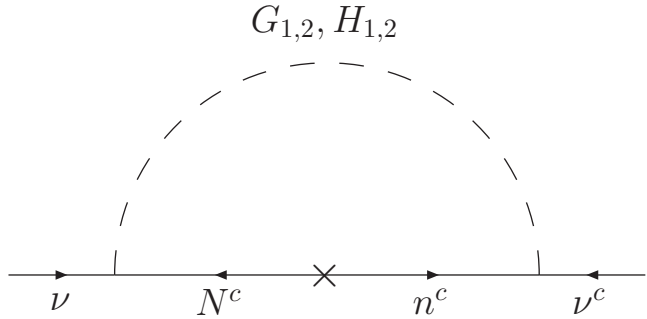
\begin{figure}[htb]
\begin{center}
\begin{picture}(480,120)(0,0)
\ArrowLine(120,10)(160,10)
\ArrowLine(240,10)(160,10)
\ArrowLine(240,10)(320,10)
\ArrowLine(360,10)(320,10)
\DashCArc(240,10)(80,0,180)9

\Text(140,0)[]{\large $\nu$}
\Text(340,0)[]{\large $\nu^c$}
\Text(200,0)[]{\large $N^c$}
\Text(280,0)[]{\large $n^c$}
\Text(240,105)[]{\large $G_{1,2},H_{1,2}$}
\Text(240,10)[]{\Large $\times$}

\end{picture}
\end{center}
\caption{One-loop generation of neutrino mass from the Goldstone bosons 
$G_{1,2}$ and Higgs bosons $H_{1,2}$.}
\end{figure}

The calculation of the $\nu \nu^c$ radiative mass is straightforward.
Using the Feynman gauge, we find that it is given by \bea
&& m_1 m_N m_{n^c} \left\{ g_N^2 \int \frac{d^4k}{(2\pi)^4} {1 \over (k^2 -
m_N^2)} {1 \over (k^2-m_{n^c}^2)}  \left( {1 \over k^2 - m_{X_1}^2} + {1
\over k^2 - m_{X_2}^2} \right) + \right. \\ && \left. {1 \over \sqrt{2} u_1
u_2} \int \frac{d^4k}{(2\pi)^4} {k^2 \over (k^2 - m_N^2)(k^2-m_{n^c}^2)}
\left( {c_1 s_1 \over k^2-m_{X_1}^2}
- {c_1 s_1 \over k^2-m_{H_1}^2} + {c_2 s_2 \over k^2-m_{X_2}^2} - {c_2 s_2
\over k^2-m_{H_2}^2} \right) \right\}. \nonumber
\eea
Note that the $H_{1,2}$ contributions are crucial in the above to make the
second integral finite.  Using
\bea
&& \int {d^4 k \over (2\pi)^4} {1 \over (k^2-a)} {1 \over (k^2-b)} {1 \over
(k^2-c)} = f(a,b,c) \nonumber \\ &=& {1 \over 16 \pi^2 i} \left[ {a \ln a
\over (a-b)(a-c)} + {b \ln b \over (b-a)(b-c)} + {c \ln c \over (c-a)(c-b)}
\right], \\ && \int {d^4 k \over (2\pi)^4} {k^2 \over (k^2-a)(k^2-b)} \left(
{1 \over k^2-c} - {1 \over k^2-d} \right) = c~f(a,b,c) - d~f(a,b,d),
\eea
we obtain
\bea
&& m_1 m_N m_{n^c} \left\{ g_N^2 [f(m_N^2,m_{n^c}^2,m_{X_1}^2) +
f(m_N^2,m_{n^c}^2,m_{X_2}^2)] \right. \nonumber \\ && \left. + {c_1 s_1
\over \sqrt{2} u_1 u_2} [m_{X_1}^2  f(m_N^2,m_{n^c}^2,m_{X_1}^2) -
 m_{H_1}^2  f(m_N^2,m_{n^c}^2,m_{H_1}^2)] \right. \nonumber \\ && \left. +
{c_2 s_2 \over \sqrt{2} u_1 u_2} [m_{X_2}^2  f(m_N^2,m_{n^c}^2,m_{X_2}^2) -
 m_{H_2}^2  f(m_N^2,m_{n^c}^2,m_{H_2}^2)] \right\}.
\eea Assuming that all heavy particles are roughly equal in mass and $g_N
\simeq g_2$, then
\beq
m_{\nu \nu^c} \simeq {g_N^2 m_1 \over 8 \pi^2} \simeq {\alpha m_1 \over 2
\pi \sin^2 \theta_W}.
\eeq
Let $m_1 \simeq 0.1$ GeV and $m_{\nu^c} \simeq 1$ TeV, then the seesaw
neutrino mass is about 0.3 eV.

Since all the particles inside the loop in Figs.~1 and 2 have odd $R$, 
the radiative Dirac neutrino mass is ``scotogenic'', i.e. caused by 
darkness.  The first such example~\cite{m06} was that of a radiative 
Majorana neutrino mass using a second scalar doublet $(\eta^+,\eta^0)$ 
which is odd under an extra $Z_2$ symmetry together with three neutral 
fermion singlets $N_{1,2,3}$, now often referred to as radiative seesaw.  
Both $\sqrt{2} \hbox{Re}(\eta^0)$ and $N_1$ have been studied as possible 
dark-matter candidates.  
In the $SU(2)_N$ model~\cite{dm11,bdmw11}, $X_1$ was studied as 
the first example of a vector-boson dark-matter candidate not from 
exotic physics. As mentioned above, this scenario can also be realized
here with radiative neutrino mass generation. However, depending on the 
region of parameter space, $H_1 $ is also a possible dark-matter candidate, 
which we consider below.

The complete Higgs potential of this model is given by~\cite{bdmw11}
\begin{eqnarray}
V &=& \mu_1^2 Tr(\phi_{13}^\dagger \phi_{13}) + \mu_2^2 \phi_2^\dagger \phi_2 
+ \mu_\chi^2 \chi \chi^\dagger + \mu_\Delta^2 Tr(\Delta^\dagger \Delta) 
+ (\mu_3^2 det \Delta + H.c.) \nonumber 
\\ &+& (\mu_{22} \tilde{\chi} \phi_{13}^\dagger \tilde{\phi}_2 + \mu_{12} 
\chi \Delta \tilde{\chi}^\dagger + \mu_{23} \tilde{\chi} \Delta {\chi}^\dagger 
+ H.c.) +  {1 \over 2} \lambda_1 [Tr(\phi_{13}^\dagger \phi_{13})]^2 + 
{1 \over 2} \lambda_2 (\phi_2^\dagger \phi_2)^2  \nonumber \\ 
&+&  {1 \over 2} \lambda_3 Tr(\phi_{13}^\dagger \phi_{13} \phi_{13}^\dagger 
\phi_{13}) + {1 \over 2} \lambda_4 (\chi \chi^\dagger)^2  +{1 \over 2} 
\lambda_5 [Tr(\Delta^\dagger \Delta)]^2 + {1 \over 4} 
\lambda_6 Tr(\Delta^\dagger \Delta - \Delta \Delta^\dagger)^2 \nonumber \\ 
&+& f_1 \chi \phi_{13}^\dagger \phi_{13} \chi^\dagger + f_2 \chi 
\tilde{\phi}_{13}^\dagger \tilde{\phi}_{13} \chi^\dagger  + f_3 \phi_2^\dagger 
\phi_{13} \phi_{13}^\dagger \phi_2 + f_4 \phi_2^\dagger \tilde{\phi}_{13} 
\tilde{\phi}_{13}^\dagger \phi_2 + f_5 (\phi_2^\dagger \phi_2)(\chi \chi^\dagger) 
\nonumber \\ &+& f_6 (\chi \chi^\dagger) 
Tr(\Delta^\dagger \Delta) + f_7 \chi (\Delta^\dagger \Delta - \Delta 
\Delta^\dagger) \chi^\dagger + f_8 (\phi_2^\dagger \phi_2) 
Tr(\Delta^\dagger \Delta) \nonumber \\ &+& f_9 
Tr(\phi_{13}^\dagger \phi_{13}) Tr(\Delta^\dagger \Delta) + f_{10} 
Tr(\phi_{13} (\Delta^\dagger \Delta - \Delta \Delta^\dagger) \phi_{13}^\dagger),
\end{eqnarray}
where
\begin{equation}
\tilde{\phi}_2 = \pmatrix{\bar{\phi}_2^0 \cr -\phi_2^-}, ~~~ \tilde{\phi}_{13} = 
\pmatrix{\phi_3^+ & -\phi_1^+ \cr - \bar{\phi}_3^0 & \bar{\phi}_1^0}, ~~~ 
\tilde{\chi} = (\bar{\chi}_2^0, - \bar{\chi}_1^0),
\end{equation}
and the $\mu_3^2, \mu_{23}$ terms break $L$ softly to $(-1)^L$.
Assuming $u_1=u_3$, and using the details provided in Ref.~\cite{bdmw11}, 
we find that $H_1 = \sqrt{2} \Delta_{2I}$ to a very good approximation, 
i.e. neglecting the terms of order $v_{1,2}/u_{1,2}$.  In that case, 
\bea
m^2_{H_1} &=& 4 \lambda_6 u_1^2 + 2 \mu_3^2 - (\mu_{12} + \mu_{23})
{u_2^2 \over 2 u_1}, \\ 
m^2_{H_2} &=& - (\mu_{12} + \mu_{23}) {8 u_1^2 + u_2^2 \over 2 u_1}.
\eea
Assuming that $H_1$ is the lightest of all particles of odd $R$, 
we now estimate its annihilation cross section in the early Universe 
and its spin-independent cross section with nuclei at underground 
experiments.

The relic density of $H_1$ is related to the $H_1 H_1$ annihilation cross 
section times the relative velocity of the two particles as they become 
non-relativistic.  This comes mainly from the contact interactions 
\beq
{1 \over 2} f_8 H_1^2 (\phi_2^+ \phi_2^- + \phi_2^0 \bar{\phi}_2^0) + 
{1 \over 2} f_9 H_1^2 (\phi_1^- \phi_1^+ + \phi_1^0 \bar{\phi}_1^0), 
\eeq
and the $\chi_{2R},\, \Delta_{1R}$ and $\Delta_{3R}$ $s-$channel
exchanges; the $t-$ and $u-$channel exchanges being 
suppressed by factors $ \sim v_i/u_j $. Taking $f_{7,10} =0,\, 
\mu_{12} = \mu_{23} (=\mu)$ to simplify the expressions, the  
$H_1 H_1 \to \phi_1 \phi_1^\dagger$ amplitude is then given by
\beq
 -i f_9
- 4 i \frac{
f_2 f_6 u_2^2 (s + 8 \lambda_6 u_1^2+ 2 \mut) + 2f_9 u_1^2
[ (s-2 \lambda_4 u_2^2)(\lambda_5 + 2 \lambda_6) + 2 f_6( f_6 u_2^2 + \mut)]
+4 \mut f_2 u_1^2 (\lambda_5 + 2 \lambda_6)}
{s^2 - 2 s (2 \lambda_5 u_1^2+ \lambda_4 u_2^2 - \mut)- 8 (f_6^2
- \lambda_4 \lambda_5) u_1^2 u_2^2
- 4 \mut(4 f_6 u_1^2 + \lambda_4 u_2^2 + 2 u_1 \mu)
} \,,
\label{eq:ampl}
\eeq
where $\mut = \mu u_2^2/u_1$, with the first term coming from the contact 
interaction. 
The corresponding amplitude for $H_1 H_1 \to \phi_2 \phi_2^\dagger$
is obtained by replacing $ f_2 \to f_5 ,\, f_9 \to f_8 $.

In order to get an estimate of the relic-density
constraints for this model, we consider two simple cases.
First suppose $ |f_6|,\,|\lambda_5|,\,|\lambda_6 | \ll 1 $:
in this case
\beq
\sigma_{\phi_1}
= \frac{|f_9|^2}{16\pi \beta s}\,,
\quad
\sigma_{\phi_2}
= \frac{|f_8|^2}{16\pi \beta s} \,,
\label{eq:cs1}
\eeq
where $ s \beta  = \sqrt{ s(s - 4  m_{H_1}^2)} $. We use 
the standard expressions~\cite{kolb},
\bea
\vevof{\sigma v} &=& \frac x{8\pi m_{H_1}^5 \, [K_2(x)]^2}
\int_{4 m_{H_1}^2}^\infty ds \, \sqrt{s} \, \left( s- 4 m_{H_1}^2 \right)
K_1\left( \frac{\sqrt{s}}T \right)
 \left(\sigma_{\phi_1}  + \sigma_{\phi_2} \right) \,,\cr
&& \cr
\Omega_{\mathrm{DM}} h^2  
&=&1.06 \times 10^9 \mathrm{GeV}^{-1}
\frac{x_f}{\sqrt{g_{*}} \, M_{Pl} \vevof{\sigma v}} \,, \quad
x_f  = \log  \left( 0.038 
\frac{\vevof{\sigma v}  M_{Pl} \, m_{H_1} }{\sqrt{g_{*} x_f }} \right) \,,
\label{eq:dm}
\eea
where $ x_f = m_{H_1}/T_f$, and $T_f $ is the freeze-out
temperature, $g_*$ the effective number of relativistic degrees
of freedom at $T=T_f$.  Taking $g_*=110.75$,
corresponding to a two-Higgs-doublet extension of the 
Standard Model, and using the experimental constraint 
$ \Omega_{DM} h^2 = 0.11 \pm 0.018 $, we find to a good approximation
\beq
\frac{m_{H_1}}{\bar f} = 4.3^{+0.4}_{-0.3} \, 
{\rm TeV} \,, \quad \bar f = \sqrt{\frac{|f_8|^2+|f_9|^2}2} \,.
\label{eq:bound.cont}
\eeq

As a second example we take the 
case $ f_6\ll1 $
and $ \mu_{3,\,12,\,23} \ll u_{1,2} $; then 
the total cross sections
equal
\beq
\sigma_{\phi_1}
= \frac{|f_9|^2}{16\pi \beta s}
\left( \frac{s+ (\rat+4) m_{H_1}^2  }{s - \rat m_{H_1}^2} \right)^2\,,
\quad
\sigma_{\phi_2}
= \frac{|f_8|^2}{16\pi \beta s}
\left( \frac{s + (\rat+4) m_{H_1}^2  }{s- \rat m_{H_1}^2} \right)^2 \,,
\qquad \rat = \frac{\lambda_5}{\lambda_6} \,,
\label{eq:cs2}
\eeq
where we take $ \rat> 1 $ to insure that $H_1 $ is the lightest dark-matter 
candidate, and also assume $ \rat< 4 $ in order to avoid complications 
associated with
resonant contributions to $ \vevof{\sigma v } $.
Using Eq.~(\ref{eq:dm}) we find 
\renewcommand{\arraystretch}{1.5}
\beq
\arrayrulecolor{blue}{
\begin{array}{|c|c|}
\hline
\rat & m_{H_1}/\bar f ~~ (\rm{TeV})\cr\hline
2	& 18.1^{+1.8}_{-1.3} \cr \hline
2.5	& 23.9^{+2.2}_{-1.8} \cr \hline
3	& 34.1^{+3.2}_{-2.5} \cr \hline
3.5	& 58.2^{+5.4}_{-4.3} \cr \hline
\end{array}}
\eeq

\noindent The factor of $ \sim 10 $ increase compared to 
Eq.~(\ref{eq:bound.cont}) is produced by a correspondingly larger
cross section from Eq.~(\ref{eq:cs1}) to Eq.~(\ref{eq:cs2}): i.e. 
by a factor of $[(\rat+8)/(\rat-4)]^2 $ in the limit $s = 4 m_{H_1}^2$.

In underground dark-matter direct-search experiments, the spin-independent 
elastic cross section for $H_1$ scattering off a nucleus of $Z$ protons 
and $A-Z$ neutrons normalized to one nucleon is given by
\beq
\sigma_0 = {1 \over \pi} \left( {m_N \over m_{H_1} + A m_N} \right)^2 
\left| {Z f_p + (A-Z) f_n \over A} \right|^2,
\eeq
where $m_N$ is the mass of a nucleon, and $f_{p,n}$ come from Higgs 
exchange~\cite{hiny11}:
\bea
{f_p \over m_p} &=& = \left( -0.075 - {0.925 (3.51)(2) \over 27} \right) 
{\sqrt{2} \over m_\phi^2} \left( {f_8 v_2^2 + f_9 v_1^2 \over v_1^2 + v_2^2} 
\right), \\ 
{f_n \over m_n} &=& = \left( -0.078 - {0.922 (3.51)(2) \over 27} \right) 
{\sqrt{2} \over m_\phi^2} \left( {f_8 v_2^2 + f_9 v_1^2 \over v_1^2 + v_2^2} 
\right).
\eea
Assuming an effective $m_\phi = 125$ GeV and using $Z=54$ and $A-Z=77$ for 
$^{131}$Xe, we find for $f_8 \simeq f_9$ with Eq.~(29) that 
$\sigma_0 < 3.2 \times 10^{-10}$ pb, which is far below the current 
upper bound of the 2011 XENON100 experiment~\cite{xenon11}.

In conclusion, we have shown how the $SU(2)_N$ model of vector-boson dark
matter allows also the one-loop radiative generation of a Dirac neutrino
mass through vector and scalar exchange.  This is the first example of such
a one-loop mechanism involving vector bosons.  We have also studied the 
scalar $H_1$ as a possible dark-matter candidate.


This research is supported in part by the U.~S.~Department of Energy under
Grant No.~DE-AC02-06CH11357.

\bibliographystyle{unsrt}

\end{document}